\documentstyle[twoside,fleqn,espcrc2,epsf]{article}

\newcommand{\AmS}{{\protect\the\textfont2
  A\kern-.1667em\lower.5ex\hbox{M}\kern-.125emS}}

\title{
\vskip -100pt
\mbox{} \hfill BI-TP 2003/25\\
\mbox{} \hfill SWAT/03/379\\
\vskip 65pt
Where is the chiral critical point in 3-flavor QCD?
\thanks{This work is supported by 
BMBF under grant No.06BI102, DFG under grant FOR 339/2-1 
and PPARC grant PPA/a/s/1999/00026.
}}

\author{ F. Karsch\rlap,
\address{Fakult\"{a}t f\"{u}r Physik, Universit\"{a}t 
Bielefeld, D-33615 Bielefeld, Germany}
C.R. Allton\rlap,\address{Department of Physics, University of 
Wales Swansea, Singleton Park, Swansea, SA2 8PP, U.K.} 
S.~Ejiri\rlap,$^{\rm a}$ S.J. Hands\rlap,$^{\rm b}$ 
O. Kaczmarek\rlap,$^{\rm a}$  E. Laermann\rlap,$^{\rm a}$ and 
C. Schmidt$^{\rm a}$}

\begin{document}

\begin{abstract}
We determine the location of the second order endpoint of the
line of first order chiral phase transition in 3-flavor QCD
at vanishing chemical potential ($\mu_q$). Using Ferrenberg-Swendsen 
reweighting for two values of the quark mass we determine the 
$\mu_q$-dependence of the transition line and locate the
chiral critical point. For both quantities we find a significant 
quark mass dependence.
\vspace{1pc}
\end{abstract}

\maketitle

\section{Introduction}
\label{sec:intro}

\noindent
Sufficiently hot and dense hadronic matter undergoes a transition to
a deconfined, chirally symmetric medium. It is expected that in QCD with 
the correct,
physically realized spectrum of quark masses this transition is a true
phase transition, related to non-analyticities of the QCD partition
function, only for specific choices of temperature ($T$) and baryon
chemical potential ($\mu_B \equiv 3 \mu_q$).
The transition then is first order for
large values of $\mu_B$, ends in a second order endpoint -- {
the chiral critical point} -- at a certain value $\mu_B^{crit}$ and is a
crossover transition, not related to singular behavior in thermodynamic
observables, for smaller values of the chemical potential (Fig.~1).

The chiral critical point is part of a critical surface in the 3-d
parameter space of degenerate light ($m_{u,d}$) and a strange ($m_s$)
quark mass as well as $\mu_B$. At $\mu_B = 0$ this surface contains
the line of second order phase transitions, which separates the region of first
order phase transitions for light quarks from a broad crossover region.

\begin{figure}[t]
\centerline{
\epsfxsize=6.0cm\epsfbox{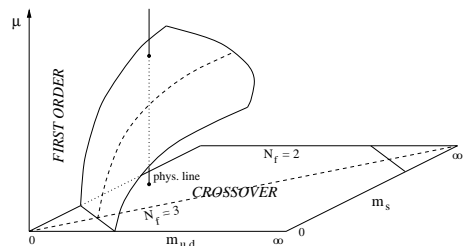}
}
\vspace*{-11mm}
\caption{Schematic plot of the critical surface in the space of
degenerate u,d-masses, the strange quark mass and the (baryon) chemical
potential. 
}
\vspace*{-3mm}
\vspace{-0.3cm}
\label{fig:surf}
\end{figure}

At $\mu_B =0$ we determine the location of the 3-flavor critical point
on this transition line. Using reweighting in $\mu_B$ we
determine the curvature of the critical surface and estimate
the location of the chiral critical point, $\mu_B^{crit}$. 
All calculations have been performed with improved staggered fermions
(p4-action) and a Symanzik-improved gauge action \cite{c2,c9}.

\begin{figure}[t]
\centerline{
\epsfxsize=6.0cm\epsfbox{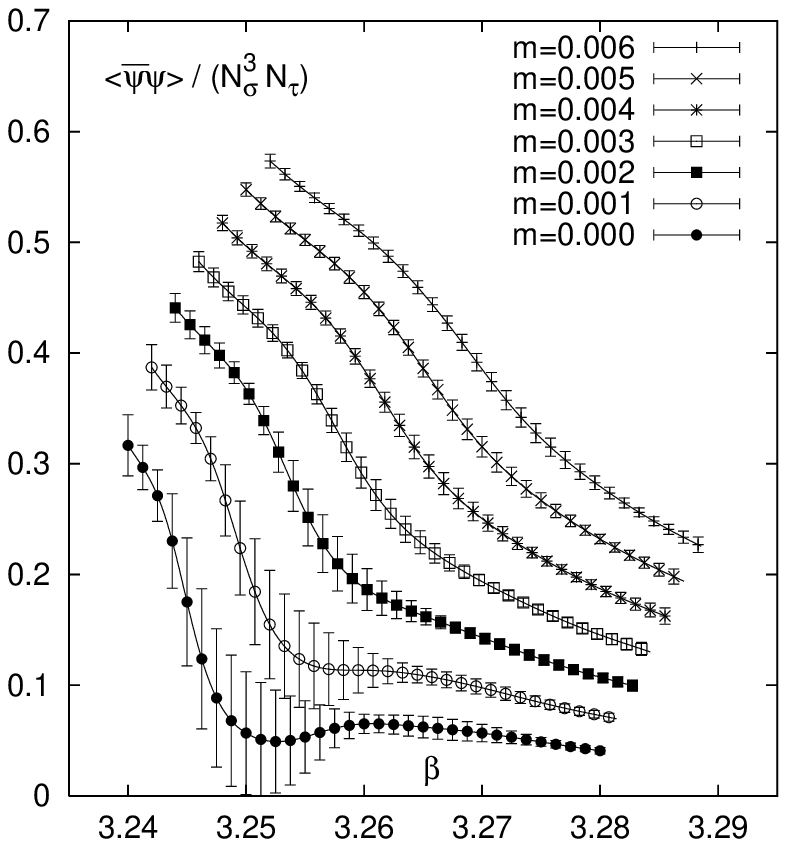}
}
\centerline{
\epsfxsize=6.0cm\epsfbox{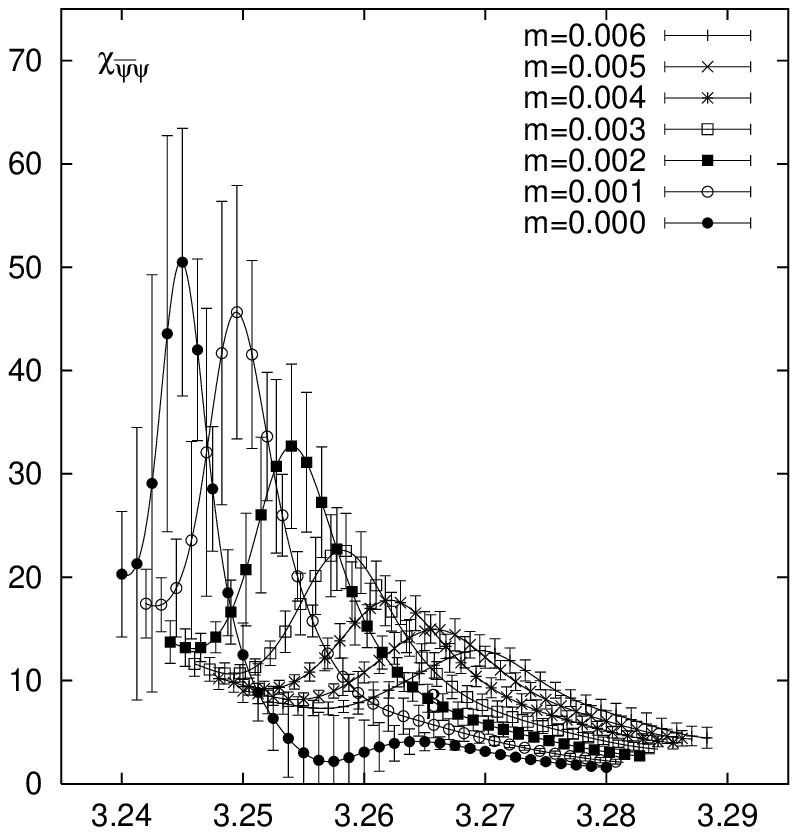}
}
\vspace*{-11mm}
\caption{Chiral condensate and chiral susceptibility at different
values of $m_q$ obtained from reweighting in $m_q$. Results
are based on data obtained on $16^3\times 4$ lattices and $m_q=0.005$
\cite{c4}. 
}
\vspace*{-6mm}
\label{fig:c2c4}
\end{figure}

\section{Critical point at $\mu = 0$}
\label{sec:critical}

\noindent
In previous studies with unimproved staggered fermions (see, for
instance \cite{c4}) the chiral critical point in 3-flavor QCD
has been located at a pion mass value of about 300~MeV. Using
Binder cumulants constructed from the chiral condensate, 
$B_4(0) = \langle (\delta \bar{\psi}\psi)^4 \rangle / \langle
(\delta \bar{\psi}\psi)^2 \rangle^2$, 
it has been verified that this point belongs to
the Ising universality class \cite{c1}. Using the knowledge of the
universality class as an input, first calculations with improved
staggered fermions (p4-action) on a $12^3\times 4$ lattice
suggested that the critical point shifts to substantially smaller
values. In \cite{c4} a preliminary value of $m_\pi^{crit}\simeq 190$~MeV
has been reported.

Our present analysis with the p4-action is based on simulations on
$12^3\times 4$ and $16^3\times 4$ with a bare quark mass of $ma=0.005$
which corresponds to a pion mass of about 170~MeV. A finite size scaling
analysis should allow to draw conclusions on the nature of the transition.
At this light pion mass we find no indication for a significant finite size
dependence of the peak height of chiral ($\chi_{\bar{\psi}\psi}$)
and Polyakov loop ($\chi_L$) susceptibilities,
\begin{equation}
\chi_{\bar{\psi}\psi} = 
\langle (\bar{\psi}\psi)^2 \rangle - \langle \bar{\psi}\psi \rangle^2 
~~,~~
\chi_L = \langle L^2 \rangle - \langle L \rangle^2 ~.~
\end{equation}
Furthermore, the Binder cumulant $B_4(0)$ 
is significantly larger than the 
3d-Ising value, $B_4^{Ising} = 1.604$.
This indicates that the chiral critical point has to lie at even
smaller values of $m_\pi$. Reweighting of both observables in the
quark mass around the simulation point at $ma=0.005$ suggests a critical 
value of $ma = 0.0007(4)$. At this value of the quark mass the Binder
cumulant is consistent with the Ising value and the peak heights in the
chiral susceptibility show a significant volume dependence which is
consistent with $\chi_{\bar{\psi}\psi}^{\rm peak} \sim V$. Some results
for the reweighted chiral condensate and $\chi_{\bar{\psi}\psi}$
obtained on the $16^3\times 4$ lattice are shown in Fig. 2. The 
rather small value found for the critical quark mass translates into a 
small value of the pion mass, which is only about half its physical
value. Such a small value is consistent with the small values 
typically found in effective model calculations and follows the trend
towards a smoother transition found in other calculations with improved
staggered fermion actions \cite{c5,c6}. It differs significantly 
from the result obtained with standard, unimproved staggered fermions. The
actual location of the chiral critical point at $\mu_B=0$ thus seems
to be strongly cut-off dependent. In summary we find, 
\begin{equation}
      m_{\pi}^{\rm crit}=\left\{\begin{array}{rl}
          290(20)\, \mbox{MeV}, &\hspace{-0.2cm} \mbox{standard action}    \\
           67(18)\, \mbox{MeV}, &\hspace{-0.2cm} \mbox{p4 improved action} \\
               47\, \mbox{MeV}, &\hspace{-0.2cm} \mbox{linear sigma model}
               \end{array} \right.
\end{equation}

\section{Reweighting in $\bf\mu_q$ at $\bf ma = 0.005$}
\label{sec:reweighting}

\vspace{4\unitlength}
\noindent
At fixed value of the quark mass we use \cite{c2} Taylor expanded
reweighting to ${\cal O}(\mu_q^2)$ of $\chi_{\bar{\psi}\psi}$ and
$B_4(0)$ to locate the chiral critical point as that point at which
$B_4$ attains the 3d-Ising value. The reweighting is done
only in the 2-flavor sector for (u,d)-quarks, {\it i.e.}
we keep $\mu_{strange}=0$.  

Reweighting
was possible only for the data on the $12^3\times 4$ lattice
as reweighting to large enough values of $\mu_q$  was prohibited
on the larger lattice due to too large fluctuations in the sign of
the fermion determinant. Even on the smaller lattice a tendency of
$B_4$ to decrease to the Ising value only sets in for $\mu_qa > 0.05$.
A linear extrapolation for $\mu a \ge 0.05$ yields
$\mu_q^{crit}a = 0.074(13)$ or
$\mu_q^{crit}/T=0.296(52)$. An estimate for the transition temperature at the
corresponding pion mass is obtained from \cite{c7} 
$T_c/\sqrt{\sigma}=0.40(1)+0.039(4)(m_\pi/\sqrt{\sigma})$. This sets the 
scale for our estimate of the critical value of the chemical potential 
in 3-flavor QCD, 
$\mu_q^{crit}=52(10)~{\rm MeV}$ at $m_\pi=170~{\rm MeV}$.
A quadratic interpolation 
to the physical pion mass value fixes the location of the chiral
critical point in 3-flavor QCD,
\begin{equation}
\mu_q^{crit} = 40(9)~{\rm MeV},~m_\pi=140~{\rm MeV}~.
\end{equation}

As expected the value,  $\mu_B^{crit} = 120~{\rm MeV}$, found here
in 3-flavor QCD is substantially smaller than the value of 
about 700~MeV estimated for (2+1)-flavor QCD \cite{c8}.

\section{The $\bf \mu$-dependence of the transition line}

\noindent
Following the prescription outlined in \cite{c1} 
the $\mu$-dependence of the transition temperature 
has been obtained using Taylor expanded reweighting of susceptibilities
to ${\cal O} (\mu_q^2)$. 
From the shift of the peak
positions one finds the critical couplings, $\beta_c(\mu_q)$, which to
leading order depend linearly
on $\mu_q^2$ and yield an estimate for the slope of $T_c(\mu_q)$:
\begin{equation}
{{\rm d}^2 T_c \over {\rm d} \mu_q^2} = {1 \over N_\tau^2 T_c (0)}
{{\rm d}^2 \beta_c \over {\rm d} \mu_q^2} / \left( a
{{\rm d} \beta_c \over {\rm d} a}\right)
\end{equation}
Comparing calculations in 3-flavor QCD for $ma=0.005$ and 0.1 we find
a significant increase in the slope of $T_c(\mu_q)$ with decreasing
$ma$.
Using the asymptotic $\beta$-function in (4) gives

\vspace{0.1cm}
\noindent
$\displaystyle{{T_c(\mu_q) \over T_c(0)} = \left\{\begin{array}{rl}
\hspace{-0.1cm}1 - 0.025(6) \left( {\mu_q \over T_c(0)} \right)^2, &\hspace{-0.2cm} 
ma = 0.1 \\
\hspace{-0.1cm}1 - 0.114(46)\left( {\mu_q \over T_c(0)} \right)^2, &\hspace{-0.2cm} 
ma = 0.005
               \end{array}
\right. }$

\vspace{0.1cm}

We note that an additional quark mass dependence is hidden here in
the transition temperature at $m_q=0$, {\it i.e.} $T_c(0)$, which is
used to normalize the transition temperature at $\mu_q\ne 0$. 
Furthermore, taking into account violations of asymptotic scaling in the
$\beta$-function will lead to a further increase of the curvature 
of $T_c(\mu_q)$. At $ma=0.01$ deviations from asymptotic scaling have
been analyzed in detail and suggest that the slope increases  
by almost a factor of 2 [5].

The transition lines for the two different quark masses together with
the estimates for the location of the chiral critical points in 3-flavor
QCD (left and right boxes) and the ($\mu_q$-independent) transition line 
in the infinite quark mass limit are shown in Fig.~3.

\begin{figure}[t]
\centerline{
\epsfxsize=6.0cm\epsfbox{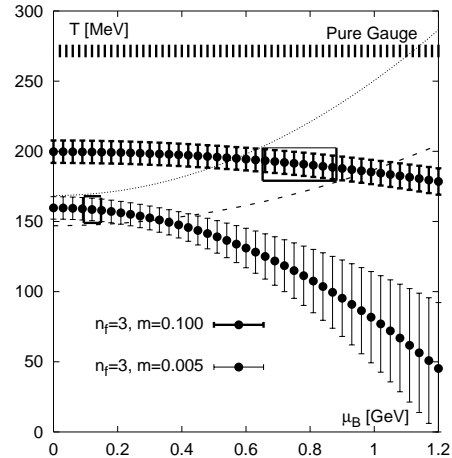}
}
\vspace*{-11mm}
\caption{The transition line for different values of the quark mass and
the quark mass dependence of the chiral critical point (boxes).
}
\vspace*{-6mm}
\label{fig:Tc2Tc3}
\end{figure}

\section{Conclusions}

At vanishing chemical potential the cut-off dependence of the chiral
critical point is large. Calculations with an improved action suggest
that the chiral transition in 3-flavor QCD becomes first order
at much smaller values of the pion mass than estimated previously.
These findings are consistent with other studies of QCD thermodynamics
performed with improved staggered fermions \cite{c5,c6}.
A similar influence of cut-off effects is found for the dependence of
the chiral critical point on the chemical potential.

\end{document}